\documentclass[twocolumn,aps,showpacs]{revtex4}
\usepackage{graphicx}
\usepackage{amsmath}
\usepackage{amsfonts}
\usepackage{amssymb}

\begin{document}

\title{\textbf{Nonlinear Barab\'asi-Albert Network}}
\author{Roberto N. Onody}
\email{onody@if.sc.usp.br}
\author{Paulo A. de Castro}
\email{pac@if.sc.usp.br}
\affiliation{Departamento de F\'{\i}sica e Inform\'atica,
Instituto de F\'{\i}sica de S\~ao Carlos, \\
Universidade de S\~ao Paulo, C.P.369, 13560-970 S\~ao Carlos-SP,
Brazil}

\begin{abstract}
In recent years there has been considerable interest in the structure and dynamics of
complex networks. One of the most studied networks is the linear Barab\'asi-Albert model. Here we investigate
the nonlinear Barab\'asi-Albert growing network. In this model, a new node connects to a vertex of degree $k$ with
a probability proportional to $k^{\alpha}$ ($\alpha$ real). Each vertex adds $m$ new edges
to the network.
We derive an analytic expression for the degree distribution $P(k)$ which is valid for all
values of $m$ and $\alpha \le 1$. In the limit $\alpha \rightarrow -\infty$ the network is homogeneous.
If $\alpha > 1$ there is a gel phase with $m$ super-connected nodes.
It is proposed a formula for the clustering coefficient which is in good agreement with numerical
simulations. The assortativity coefficient $r$ is determined and it is shown that the
nonlinear Barab\'asi-Albert network is assortative (disassortative) if $\alpha < 1$ ($\alpha > 1$)
and no assortative only when $\alpha = 1$. In the limit $\alpha \rightarrow -\infty$ the
assortativity coefficient can be exactly calculated. We find $r=\frac{7}{13}$ when $m=2$.
Finally, the minimum average shortest path length $l_{min}$ is numerically evaluated.
Increasing the network size, $l_{min}$ diverges for $\alpha \le 1$ and it is equal to 1
when $\alpha > 1$.


\end{abstract}
\maketitle

\section{Introduction}

In the late 1990s, the researches on complex networks knew an
explosive development \cite{new0,dm,br}. The impulse came from the convergence of interests
between different sectors like physics, biology, technology and sociology turning the study
of complex networks into a new interdisciplinary field.

A network is a set of vertices or nodes provided with some rule to
connect them by edges. With such a simple definition, it is not a
surprise that it has found so much utility in so many areas.
Following the classification introduced by Newman \cite{new0}, one
can divide real networks into four types: social, biological,
technological and informational.

Social networks are composed by interacting people with some
pattern of contacts like friendship, business or sexual partners.
One of the most popular works in social webs was carried out by
Milgram \cite{mil} who first arrived to the concept of "six
degrees of separation" and small-world. Social networks also
include problems like the collaboration network of film actors
\cite{ws,ah} or co-authorship among academics \cite{betal}.

A number of problems can be mapped into biological networks. Some
very good examples are: the 282-neuron neural network of the
nematode {\it C. Elegans} \cite{wstb}, blood vessels and vascular
networks \cite{bmr,wbe}, food webs in an ecosystem with species
living in a prey-predator scheme \cite{pimm,cgar}, network of metabolic
pathways \cite{jnb} and the genetic regulatory network for the expression of
a gene \cite{smm}.

Technological networks are those built by man in his arduous struggle for
progress and welfare. Perhaps, the best known example is the
electric power grid (high-voltage three-phase transmission lines
network) \cite{ws,asbs}. This category also includes the networks of
airline routes \cite{asbs}, railways \cite{sen} and internet structures \cite{fff}.

The fourth type is the information network. One of the oldest
example of this kind is the network of citations between academic papers
\cite{pri}. This network is certainly influenced by social
relationships and it may be somehow contaminated since putting a
citation in a paper does not mean that the author had actually
read it \cite{sim}. Another important example is the World Wide
Web, which is the network of informations between Web pages
\cite{baje}. It is a directed network with different
power laws for the in and out degree distributions.

Motivated by such a number of applications, a myriad of
theoretical models were proposed aiming to reproduce or to
describe real-world networks. In many of these models, the degree
distribution is power law and the corresponding network is said
scale-free \cite{ba,dm}. A good number of real-world networks
are of this kind.

One widely studied scale-free network is the linear
Barab\'asi-Albert model (BA) \cite{ba,baj}. In its general form,
that we call the nonlinear Barab\'asi-Albert model (NBA), the
network is constructed as follows. A new vertex connects to
another (already existing) vertex $i$ with probability $\Pi
(k_{i}) = k_{i}^{\alpha}/\sum_{j} k_{j}^{\alpha}$, where $k_{i}$
is the degree, i. e., the number of edges connected to the vertex
$i$ and $\alpha$ is a real number. Each new vertex adds $m$ new
edges to the network. In the linear case ($\alpha = 1$), the model
has many desired characteristics like to be a scale-free network
or to have the small-world effect, but it also has some other
features which are not so coveted: it is no assortative and the
clustering goes to zero with the increasing size of the network.
A network is said assortative if it connects preferentially nodes
with almost the same degree. It is something like a social stratification
by the incomes.
If we think of social networks, assortative mixing and clustering
are both yearned qualities \cite{np}.

In this paper, we investigate the NBA (the model, not the basketball association)
by determining some important quantities like the degree distribution, the
clustering coefficient $C$, the assortativity coefficient $r$ and the average shortest
path length $l$. This is done in the whole space of the parameters $m$ and $\alpha$.
We obtain some interesting analytical as well as numerical results. Writing the
master equation for the number of vertices with degree $k$, we derive an expression
for the degree distribution $P(k)$. If $\alpha > 1$, the gel phase has $m$ super-
connected sites. For $\alpha \rightarrow -\infty$ the NBA becomes a homogeneous
lattice with all sites having connectivity $2m$. Through a simple change of
stochastic variables, we rederive the known stretched exponential form of $P(k)$ in the
mean field approximation. It is proposed a formula for the clustering coefficient which
was verified by numerical simulations. The behavior of the assortativity coefficient is
numerically evaluated. The NBA is assortative (disassortative) if $\alpha < 1$
($\alpha > 1$) and no assortative only when $\alpha = 1$. In the limit $\alpha \rightarrow
-\infty $, the assortativity coefficient can be calculated exactly. We find
$r=\frac{7}{13}$ when $m=2$. By numerical simulations, the minimum average shortest
path length $l_{min}$ is determined. As the size of the network increases, $l_{min}$
diverges if $\alpha \le 1$ and it is equal to $1$ otherwise.

\section{Degree distribution using the master equation}

The probability $P(k)$ that a randomly chosen vertex has degree $k$ was obtained in
references \cite{krl,kr} for the case $m=1$. Here we derive this probability
for any value of $m$.
For a network with $N$ vertices, the master equation reads

%------------------------------------ equation  -------------------------------
\begin{equation}
\frac{dN(k)}{dt} = \frac{m}{M_{\alpha}} [(k-1)^{\alpha} N(k-1) -
k^{\alpha} N(k) ] + \delta_{k,m} \; ,
\end{equation}
%------------------------------------------------------------------------------
where $N(k)$ is the number of vertices with degree $k$ at the time $t$ and
$M_{\alpha}=\sum_{k=m}^{N} k^{\alpha} N(k)$ gives the proper normalization.
The first (second) term accounts for the process in which a vertex with $k-1$
($k$) links is connected to the new vertex. For the $m$ possibilities,
this happens with probability $ m (k-1)^{\alpha} / M_{\alpha}$ ($ m k^{\alpha}
/ M_{\alpha}$). The last term accounts for the continuous introduction of new vertices.

If $\alpha \le 1$, we expect $M_{\alpha}$ to scale linearly in the long-time
limit, i.e., $M_{\alpha}=\mu t$ ($t \sim N = \sum_{k} N(k)$).
Writing $N(k) = t P(k)$, where $P(k)$ is the stationary degree distribution,
and substituting it into the equation above we get for $k = m$

%------------------------------------ equation  -------------------------------
\begin{equation}
P(m)=\frac{\mu}{\mu+m^{1+\alpha}}
\end{equation}
%------------------------------------------------------------------------------
and the recursive relation

%------------------------------------ equation  -------------------------------
\begin{equation}
P(k)=\frac{m(k-1)^{\alpha}}{\mu+mk^{\alpha}}P(k-1), \; \; \; k \ge m+1
\end{equation}
%------------------------------------------------------------------------------
which has the solution

%------------------------------------ equation  -------------------------------
\begin{equation}
P(k)=\frac{\mu}{mk^{\alpha}}\prod_{j=m}^{k}(1+\frac{\mu}{mj^{\alpha}})^{-1} \; .
\end{equation}
%------------------------------------------------------------------------------

We can now establish the dependence of the amplitude $\mu$ on $\alpha$. Using the
relation $ \mu = \sum_{k \ge m} k^{\alpha}P(k)$ together with the equation above
we obtain the intrinsic relation

%------------------------------------ equation  -------------------------------
\begin{equation}
\mu=m^{\alpha}[1-m+\sum_{k=m+1}^{\infty}\prod_{j=m+1}^{k}(1+\frac{\mu}{mj^{\alpha}})^{-1}] \; .
\end{equation}
%------------------------------------------------------------------------------

If $\alpha < 0$, the derived expressions are still valid but the NBA model looses
its usual preferential attachment or capitalistic interpretation.
With $\alpha$ negative, the new vertex connects to those poorest vertices in a say socialistic
way. If $\alpha=-\|\alpha\|$ with $\|\alpha\| \gg 1$
and assuming the connectivity $k$ to be a continuous
variable, the degree distribution exhibits a maximum at
$k_{max}=[\frac{m\|\alpha\|}{\mu}]^{\frac{1}{\|\alpha\|}}$. The degree
distribution expanded around this maximum has the gaussian form

%------------------------------------ equation  -------------------------------
\begin{equation}
P(k) \sim \|\alpha\| \exp{(-\frac{\|\alpha\|^{2}}{2}}(k-k_{max})^{2})  \; .
\end{equation}
%------------------------------------------------------------------------------

In the limit $ \|\alpha\| \rightarrow \infty$ it approaches the Dirac Delta function
$\delta(k-k_{max})$. The condition $\sum_{k} k P(k)=2m$ fixes $k_{max}=2m$.
The network is now homogeneous with all vertices having connectivity $2m$.

If $\alpha > 1$ and $m=1$, the asymptotic behavior of $M_{\alpha}$ is
$t^{\alpha}$ \cite{krl} and there arise one super-connected node to which almost every other
vertex is connected. When $m > 1$, we found $m$ super-connected nodes. This result was confirmed
exhaustively by our numerical simulations.
For large enough networks, the degree histogram always shows (with no exceptions)
$m$ super-connected nodes.
The probability {\it B}
that the $m$ initial vertices are connected to all
other remaining sites of the network can be calculated. In the discrete time version process,
let $s=t-1$ ($t=N-m$). When the $(m+2)$-th vertex is being aggregated,
the probability that it will be connected to all $m$ initial sites is
$\theta(s=1)=\prod_{j=1}^{m}\frac{j}{j+m^{\alpha}}$. In general, we have

%------------------------------------ equation  -------------------------------
\begin{equation}
\theta(s)=\prod_{j=1}^{m}\frac{1}{1+\frac{s^{1-\alpha} m^{\alpha}}{j}}  \; ,
\end{equation}
%------------------------------------------------------------------------------
and

%------------------------------------ equation  -------------------------------
\begin{equation}
{\it B}=\prod_{s=1}^{\infty}\theta(s)  \; .
\end{equation}
%------------------------------------------------------------------------------

Clearly, $B=0$ when $\alpha \le 2$ and $B > 0$ otherwise,
reproducing the results obtained for $m=1$ \cite{krl}. Thus the threshold
$\alpha=2$ does not depend on $m$.

\section{Degree distribution in the mean field approximation}

In the mean field context, the degree distribution was obtained
for the linear BA network \cite{baj} and for the NBA model
with link fluctuations \cite{lly}. Here we rederive the results
using a very simple and straightforward change of stochastic variable.

Assuming the connectivity $k_{i}$ to be a continuous variable defined on a
site $i$ of a network with $N$ vertices, one can write

%------------------------------------ equation  -------------------------------
\begin{equation}
\frac{\partial k_{i}}{\partial t}=m \frac{k_{i}^{\alpha}}{\sum_{j=1}^{N} k_{j}^{\alpha}}  \; .
\end{equation}
%------------------------------------------------------------------------------

As in the previous section, if $\alpha \le 1$ we can assume
$ \sum_{j=1}^{N} k_{j}^{\alpha}=\mu N \sim \mu t$. Substituting this into the
expression above and integrating it, we get

%------------------------------------ equation  -------------------------------
\begin{equation}
k_{i}=m[1+\frac{m^{\alpha}(1-\alpha)}{\mu}\ln(\frac{t}{t_{i}})]^\frac{1}{1-\alpha}  \; ,
\end{equation}
%------------------------------------------------------------------------------
where $t_{i}$ is the time when the site $i$ was incorporated to the network and
the initial condition $k_{i}(t_{i})=m$ was used. Now, if one vertex is to be picked
randomly then the probability that the chosen site is $i$ is $P(t_{i})=\frac{1}{N}$,
so the equation above can be seen as a change of stochastic variables
$t_{i} \rightarrow k_{i}$ and the respective distributions are related by
$P(k)=\int \delta(k - k_{i}(t_{i})) P(t_{i}) dt_{i}$. The result is

%------------------------------------ equation  -------------------------------
\begin{equation}
P(k)=\frac{\mu}{m k^\alpha}\exp\{-\frac{\mu}{m^\alpha(1-\alpha)}[(\frac{k}{m})^{1-\alpha}-1]\}  \; ,
\end{equation}
%------------------------------------------------------------------------------
which is the asymptotic behavior of the exact $P(k)$ given by Eq.(4)
(see also \cite{lly}). We will later use this mean field result to derive
the clustering coefficient in the next section.

If the same procedure is applied to the BA with the modified preferential attachment

%------------------------------------ equation  -------------------------------
\begin{equation}
\frac{\partial k_{i}}{\partial t}=m \frac{(k_{i}+\lambda)}{\sum_{j=1}^{N}(k_{j}+\lambda)}
  \; , \; \; \lambda > -m \; ,
\end{equation}
%------------------------------------------------------------------------------
then the degree distribution is given by

%------------------------------------ equation  -------------------------------
\begin{equation}
P(k)=(2+\frac{\lambda}{m})(\frac{k+\lambda}{m+\lambda})^{-3-\frac{\lambda}{m}} \; ,
\end{equation}
%------------------------------------------------------------------------------
which, in the asymptotic limit $k \gg 1$, is a power law
$k^{-3-\frac{\lambda}{m}}$ with the exponent tuned by the parameter $\lambda$.

\section{Clustering Coefficient}

The clustering coefficient is a transitivity property of the network. If
a vertex $1$ is connected with vertex $2$ and the vertex $2$ with
the vertex $3$ then there is a high probability that the vertex
$1$ is also connected to $3$. In social networks, this can be
easily interpreted as the fact that the friend of your friend is
likely to be your friend. The clustering coefficient of order $x$
of a site $i$, $C_{i}(x)$, is defined as the probability that there is a
distance of length $x$ (measured without passing through $i$)
between two nearest neighbors of the site $i$
\cite{fhjs}. It is given by

%------------------------------------ equation  -------------------------------
\begin{equation}
C_{i}(x)=\frac{2 y_{i}}{k_{i} (k_{i}-1)} \; ,
\end{equation}
%------------------------------------------------------------------------------
where $y_{i}$ is the number of such $x$ distances and $k_{i}$ is the degree of the
vertex $i$. For all vertices of the network, the average clustering coefficient is
$C(x)=\frac{\sum_{i=1}^{N} C_{i}(x)}{N}$.
In this paper, we will only treat the order $x=1$, so we drop the index $x$ and use
simply $C$ as being the average clustering coefficient. Of course, when $m=1$, $C$ is
always equal to zero (independently of the value of $\alpha$) since the graph is a tree like.

We can derive an analytic expression for $C$ in the mean field approximation. Let
$P(j \rightarrow i)$ be the probability that at the time $t=j$
the vertex $j$ is connected to the vertex $i$. This means that
$P(j \rightarrow i)=m \frac{k_{i}^{\alpha}(j)}{\sum_{l=1}^{N} k_{l}^{\alpha}(j)}$, where
$k_{i}(j)$ is the degree of the vertex $i$ at the time $j$. For $\alpha \le 1$,
the denominator of this probability is equal to $\mu j$ and using Eq.(10) we get

%------------------------------------ equation  -------------------------------
\begin{equation}
P(j \rightarrow i)=\frac{m^{1+\alpha}[1+\frac{m^{\alpha}(1-\alpha)}{\mu}\ln(\frac{j}{i})]
^\frac{\alpha}{1-\alpha}}{\mu j}  \; \; j > i .
\end{equation}
%------------------------------------------------------------------------------

When $\alpha$ is equal to one, the expression above reduces to
$\frac{m}{2}(ij)^{-1/2}$ which was first obtained by Klemm and
Egu\'{\i}luz \cite{ke}.

To find the clustering coefficient $C$, care must be taken
with the ordering $ j > i$. They cannot be interchanged.
If $i$ and $j$ are two nearest neighbors vertices of the
node $l$ and we assume $i < j$, then the contributions come from 3 regimes:
$ l < i < j$, $i < l < j$ and $i < j < l$. In this way, the average clustering
coefficient can be written

%------------------------------------ equation  -------------------------------
\begin{eqnarray}
C = \frac{1}{N}[\sum_{j=3}^{N} \sum_{i=2}^{j-1} \sum_{l=1}^{i-1}
\frac{P(i \rightarrow l) P(j \rightarrow l) P(j \rightarrow i)}{n_{l}(N)}+ \notag \\
\sum_{j=3}^{N} \sum_{l=2}^{j-1} \sum_{i=1}^{l-1}
\frac{P(l \rightarrow i) P(j \rightarrow l) P(j \rightarrow i)}{n_{l}(N)}+ \notag \\
\sum_{l=3}^{N} \sum_{j=2}^{l-1} \sum_{i=1}^{j-1}
\frac{P(l \rightarrow i) P(l \rightarrow j) P(j \rightarrow i)}{n_{l}(N)}] \; ,
\end{eqnarray}
%------------------------------------------------------------------------------
where $n_{l}(N) = \frac{k_{l}(N)[k_{l}(N)-1]}{2}$ is the total number of pairs of
neighbors that the node $l$ has at the time (or size) $N$.

For all values of $ \alpha \le 1$, the clustering coefficient goes
to zero with $N$. There are two particular cases in which the asymptotic behavior of $C$
is known exactly: when $\alpha = 0$ (random graph) $C \propto \frac{1}{N}$ and
$\alpha = 1$, $C(N) \propto N^{-1} (\ln N)^{2}$ \cite{ke}. The latter form is
more consistent with the numerical simulations results than an earlier theoretical
prediction ($C(N) \propto N^{-0.75}$ \cite{ab}).

We simulate the nonlinear NBA model in networks with sizes varying from $N=100$ up to
$N=102,400$ and $m=2$. For each size, the average clustering coefficient was determined
and then average it over
$20$ to $24,000$ independent runs. In the Fig.1, we plot the dependence of $C$ with $\alpha$
for two network sizes $N=102,400$ and $N=25,600$.

\begin{figure}[htbp!]
\begin{center}
\includegraphics[width=8cm]{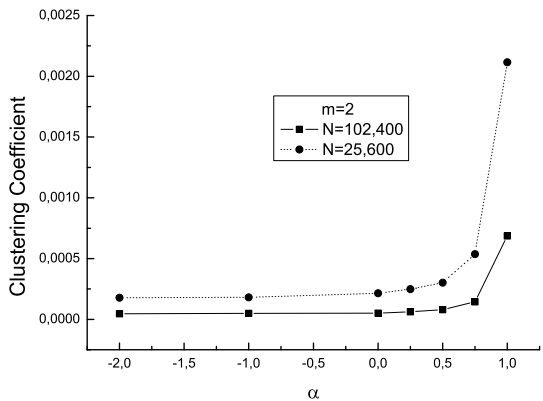}
\end{center}
\caption{The average clustering coefficient C versus the exponent $\alpha$ for two
different network sizes $N$.}%
\label{Figure1}%
\end{figure}

In the Figure 2, we compare the results of
the analytical expression (eq.(16)) with those of the simulated NBA model.

For $\alpha > 1$, the clustering coefficient rapidly approaches the value $1$ as we increase
the size $N$. This result comes from the fact
that the $m$ super-connected nodes are also inter-connected. Thus, when a new vertex is added to the network,
the strong preferential dynamics (alpha > 1) forces it to connect with the $m$ super-connected
nodes, leading to a clustering coefficient value close to one.

\begin{figure}[htbp!]
\begin{center}
\includegraphics[width=8cm]{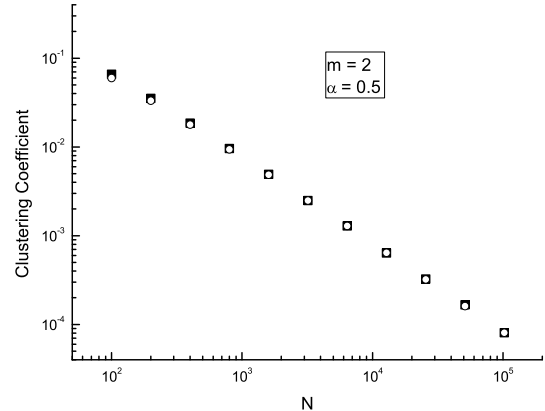}
\end{center}
\caption{Comparison between numerical simulation and theory. Black squares are the simulated points and
the open circles correspond to the proposed analytical formula}%
\label{Figure2}%
\end{figure}

\section{Mixing Patterns}

One important aspect of the networks is how the vertices are linked. If there is a tendency
to connect vertices with almost the same degree, we call it an assortative mixing. When the
links favor nodes of very different degrees the system is said disassortative.
The neutral situation is named no assortative \cite{new1,new2}. To measure the assortative
mixing level, we slightly change the definitions given by Callaway {\it et al.} \cite{call}. Let
$ e_{jk}$ be the joint probability distribution that a randomly chosen edge has degrees $j$ and
$k$ at either end. It obeys the sum rules

%------------------------------------ equation  -------------------------------
\begin{equation}
\sum_{jk} e_{jk} = 1, \;\;\; \sum_{j} e_{jk} = q_{k}.
\end{equation}
%------------------------------------------------------------------------------

Originally, $j$ and $k$ were defined as the remaining degrees - the number of edges leaving the
vertex other than the one we arrived along \cite{call}. We can define the assortativity
coefficient $r$ as \cite{new1,new2}

%------------------------------------ equation  -------------------------------
\begin{equation}
r = \frac{1}{\sigma^{2}_{q}} \sum_{jk} jk(e_{jk}- q_{j} q_{k}),
\end{equation}
%------------------------------------------------------------------------------
where $\sigma^{2}_{q} = \sum_{k} k^{2} q_{k} -[\sum_{k} k q_{k}]^{2} $ is the variance
of the distribution $q_{k}$. The quantity $r$ lies in the interval $-1 \le r \le 1$. It
has the value $+1$ ($-1$) if the system is perfectly assortative (disassortative) and it is zero
in the no assortative case.

\begin{figure}[htbp!]
\begin{center}
\includegraphics[width=8cm]{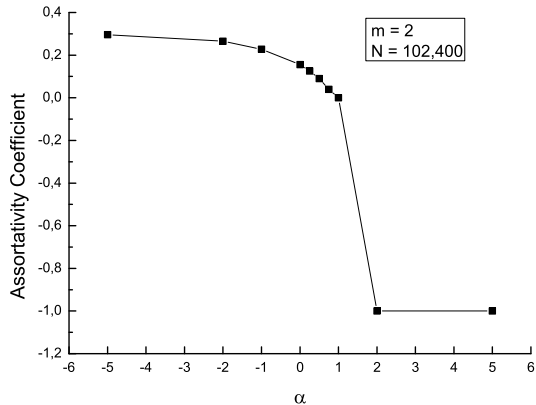}
\end{center}
\caption{The dependence of the assortativity coefficient r with the exponent $\alpha$.}%
\label{Figure3}%
\end{figure}

In our simulations, we determine $r$ and average it in a number of $20$ to $24,000$ runs depending
on the network size $N$. In the Figure 3 we plot the assortative coefficient $r$
against the exponent $\alpha$. In the limit of an infinite lattice, $r=0$ if $\alpha=1$,
a result already known \cite{new1}. Our numerical data show that this is the unique case.
The NBA model is assortative ($r > 0$) for $\alpha < 1$ and disassortative
($r < 0$) when $\alpha > 1$.

In the limit $\alpha \rightarrow -\infty$ the assortativity coefficient can be analytically calculated.
For example, if $m=2$ then the number of vertices $N(k)$ with degree $k$ is
$N(2)=1$, $N(3)=6$, $N(4)=N-7$ and zero for $k \ge 5$. The mean degree $<k>$ is equal to $4-\frac{8}{N}$. The
joint probability distribution $e_{ij}$ ($i, j = 2,3,4$) can be easily evaluated

%------------------------------------ equation  -------------------------------
\[ e_{ij} = \frac{1}{N <k>} \left( \begin{array}{ccc}
0 & 1 & 1 \\
1 & 10 & 7 \\
1 & 7 & 4N-36
\end{array} \right), \]
%------------------------------------------------------------------------------

Using the definition (Eq. 18) and the results above one can get straightforwardly
the exact value $r = \frac{7}{13}$.

\section{Average shortest path length}

With all edges of a network having unit length, the distance between two vertices
of that network is the size of the shortest path length between them (geodesic).
When we take into account all possible pairs of vertices, the mean value corresponds
to what is known as the average shortest path length $l$ (sometimes also called
the diameter). It is a fundamental concept in networks studies and very important in the
field of communications and computer's web where routing and searching are common tasks.

To estimate $l$ we evaluate $l_{min}$ which is defined as the average
distance of all vertices on the graph measured from the vertex with the
highest degree in the network (if there is more than one, we
choose one by chance). The average shortest path length $l$ is
restricted to the interval \cite{ch}

%------------------------------------ equation  -------------------------------
\begin{equation}
l_{min} \le l \le 2l_{min}.
\end{equation}
%------------------------------------------------------------------------------

We determine $l_{min}$ for several values of $\alpha$ and $m$ by simulating the
nonlinear Barab\'asi-Albert in networks with sizes
varying from $N=100$ up to $N=102,400$ and averaging it over $20$ to $24,000$ independent
realizations. Figure 4 shows how $l_{min}$ depends on $\alpha$ for two different networks sizes.
In the limit $N \rightarrow \infty$, $l_{min}$ is infinite for $\alpha \le 1$ and equal to $1$
when $\alpha > 1$.

\begin{figure}[htbp!]
\begin{center}
\includegraphics[width=8cm]{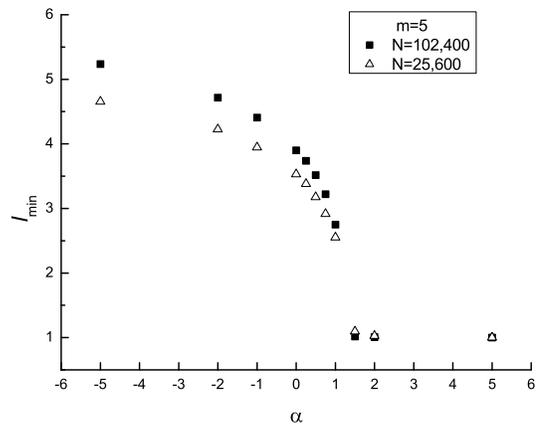}
\end{center}
\caption{The minimum average shortest path length $l_{min}$ versus the exponent $\alpha$
for two different network sizes.}%
\label{Figure4}%
\end{figure}

Let us now
discuss the dependence of $l_{min}$ with $N$. We conjecture that both $l$ and
$l_{min}$ scale with $N$ in the same way.
It is well known that random networks, such as Erd\"os-R\'enyi
networks \cite{er1,er2}, or partially random like the small-world
networks \cite{ws}, have an average shortest path length
scaling as $l \sim \ln (N)$. For scale-free networks having a
degree distribution $P(k) \sim k^{-\lambda}$, it was proved
that $l \sim \ln(\ln(N))$ ($l \sim \ln(N)$) if $ 2 < \lambda < 3$
($\lambda > 3$) \cite{ch},
i. e., they form an ultra-small-world (small-world). In the frontier value
$\lambda = 3$, the expected dependence is $l \sim \frac{\ln(N)}{\ln(\ln(N))}$ \cite{ch}.

In the mean field approximation, the linear Barab\'asi-Albert
model ($\alpha=1$) has an exponent $\lambda=3$ which is
independent of $m$. However, in the real linear Barab\'asi-Albert
model, simulated in networks with a few millions nodes, the
exponent $\lambda$ is actually less then $3$ \cite{baj}. Through a
careful analysis of our numerical data, we got the exponent
$\lambda=2.91 \pm 0.03$. This give us an opportunity to check two
theoretical predictions: if $m = 1$, the linear BA is a graph tree
and the average shortest path length scales as $l \sim \ln(N)$
\cite{br}; if $m \ge 2$, the expected dependence is $l \sim
\ln(\ln(N))$ \cite{ch} once actually $\lambda < 3$.
Our numerical results corroborate
these theoretical previsions as can be seen in the Fig. 5.

\begin{figure}[htbp!]
\begin{center}
\includegraphics[width=7.3cm]{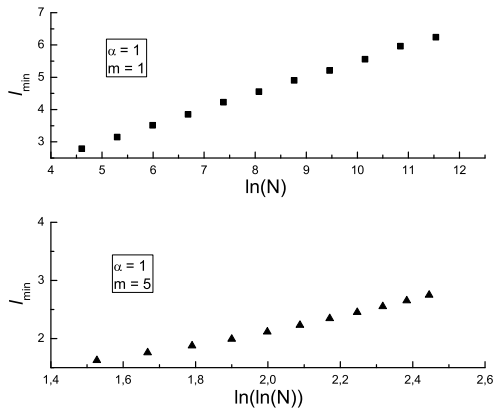}
\end{center}
\caption{The linear BA model ($\alpha=1$). Plot of the minimum average shortest path
length $l_{min}$ versus $\ln(N)$ and $\ln(\ln(N))$ for $m=1$ and $m=5$, respectively.}%
\label{Figure5}%
\end{figure}

\section{Conclusions}

We studied many properties of the nonlinear Barab\'asi-Albert network.
Using the master equation, we derive an analytic expression for the
degree distribution $P(k)$ which holds for all integer $m \ge 1$ and real
$\alpha \le 1$. In particular, the situation $\alpha < 0$ was focused
for the first time. This sector had been neglected in most of the previous
works. We claim that it might be useful in some social networks (since
the system is highly assortative there) or in the study of the crossover from the small-world
regime, where the shortest path length grows up logarithmically with the network size,
to the large-world regime, where it grows up faster with some power
of the network size. In the limit $\alpha \rightarrow -\infty$, the network is
homogeneous with connectivity $2m$. If $\alpha > 1$, there are $m$ super-connected vertices
in the gel phase. The probability that the $m$ initial vertices are connected to all the other
sites is non null only when $\alpha \ge 2$ and this threshold $\alpha=2$ does not depend on $m$.

We proposed an analytic formula for the average clustering coefficient $C$. Its validity
was verified by numerical simulations. For any fixed network size,
$C$ is a monotonically increasing function of $\alpha$. If $N \rightarrow \infty$,
the clustering coefficient falls to zero for all $\alpha \le 1$ and rapidly approaches its maximum
value $C=1$ when $\alpha > 1$.

The mixing patterns of the NBA model were determined by the assortativity coefficient $r$.
If $\alpha < 1$, the assortativity coefficient $r$ increases with $N$ and converges asymptotically
to some real positive value which is smaller than one;
if $\alpha = 1$, $r$ decreases with the size $N$ and goes to zero in the
limit of an infinite network; if $\alpha > 1$, $r$ diminishes with the size but now converges to
the value $-1$. In other words, NBA is assortative (disassortative) if $\alpha < 1$ ($\alpha > 1$)
and no assortative only when $\alpha = 1$. In the limit $\alpha \rightarrow -\infty$ the
assortativity coefficients can be exactly calculated. We find $r=\frac{7}{13}$ when $m=2$.

Through the minimum average shortest path length $l_{min}$ we estimate the diameter of the NBA. For
a fixed network size, $l_{min}$ is a monotonically decreasing function of $\alpha$.
It goes to infinite if $\alpha \le 1$ and is equal to 1 when $\alpha > 1$. In the particular (linear) case
$\alpha=1$, $l_{min} \sim \ln(N)$ if $m=1$ and $l_{min} \sim \ln(\ln(N))$ when $m \ge 2$.

This work was supported in part by the CNPq (Conselho Nacional de
Desenvolvimento Cient\'{\i}fico e
Tecnol\'ogico) and by the FAPESP (Funda\c c\~ao de Amparo a Pesquisa do
Estado de S\~ao Paulo).

\end{document}